\documentclass[10pt]{article}

\usepackage{latexsym} 
\usepackage{theorem} 

\newcommand{\forsub}[1]{\ifthenelse{\equal{\version}{sub}}{#1}{}}
\newcommand{\forfull}[1]{\ifthenelse{\equal{\version}{full}}{#1}{}}
\newcommand{\forfinal}[1]{\ifthenelse{\equal{\version}{final}}{#1}{}}

\newcommand{\comment}[1]{}

\newcommand{\home}{\raisebox{-1mm}{\~}}

\newcommand{\ie}{{\em i.e. }}

\newcommand{\lex}{_{lex}}

\newcommand{\mul}{_{mul}}

\newcommand{\stat}[1][f]{_{stat_{#1}}}
\newcommand{\state}[1][f]{_{stat_{#1}}^\simeq}
\newcommand{\statt}[1][f]{_{stat_{#1}}^>}

\renewcommand{\a}{\rightarrow}

\newcommand{\ad}{\downarrow}

\newcommand{\au}{\uparrow}
\renewcommand{\to}{\mapsto}

\newcommand{\ps}[1]{{\langle #1\rangle}}
\newcommand{\dps}[1]{\ps{\!\ps{#1}\!}}

\newcommand{\all}{\forall}

\newcommand{\sle}{\subseteq}

\newcommand{\tle}{\unlhd}
\newcommand{\tge}{\unrhd}
\newcommand{\tlt}{\lhd}
\newcommand{\tgt}{\rhd}

\newcommand{\cge}{\succeq}
\newcommand{\clt}{\prec}
\newcommand{\cgt}{\succ}

\newcommand{\h}{\widehat}

\renewcommand{\u}{\underline}

\renewcommand{\b}{\beta}

\renewcommand{\l}{\lambda}
\renewcommand{\L}{\Lambda}
\newcommand{\e}{\eta}
\newcommand{\s}{\sigma}

\renewcommand{\t}{\theta}

\newcommand{\cA}{{\cal A}}
\newcommand{\cB}{{\cal B}}

\newcommand{\cF}{{\cal F}}

\newcommand{\cH}{{\cal H}}
\newcommand{\cI}{{\cal I}}

\newcommand{\cR}{{\cal R}}

\newcommand{\cX}{{\cal X}}

\newcommand{\cZ}{{\cal Z}}

\newcommand{\beginappendixes}{\setcounter{section}{0}%
  \renewcommand{\thesection}{\Alph{section}}}
\newcommand{\append}[1]{\newpage\refstepcounter{section}%
  \setcounter{counter}{0}%
  \renewcommand{\thecounter}{\thesection.\arabic{counter}}%
  \section*{Appendix \thesection: #1}}

\newcounter{counter}

{\theorembodyfont{\rmfamily}
  \newtheorem{dfn}[counter]{Definition}
  \newtheorem{lem}[counter]{Lemma}
  \newtheorem{thm}[counter]{Theorem}
  \newtheorem{cor}[counter]{Corollary}
  
  \newtheorem{prop}[counter]{Property}

}

\newenvironment{prf}
  {Proof.}
  {}

\newcommand{\setlst}{\parsep=0cm\topsep=0cm\itemsep=0cm%
  \labelsep=1mm\itemindent=0mm}
\newenvironment{lstx}[2]
  {\begin{list}{#1}{\setlst#2}}
  {\end{list}}
\newenvironment{lst}[1]
  {\begin{list}{#1}{\setlst\labelsep=1mm\itemindent=-2mm}}
  {\end{list}}

\newenvironment{enumx}[1]
  {\begin{lstx}{}{\usecounter{enumi}%
     \renewcommand{\makelabel}{(\arabic{enumi})}%
     #1}}
  {\end{lstx}}
\newenvironment{enum}
  {\begin{enumx}{\labelsep=1mm\itemindent=-2mm}}
  {\end{enumx}}
\newenvironment{prfenum}
  {\begin{enumx}{\labelsep=-1mm\itemindent=3.3mm}}
  {\end{enumx}}

\usepackage{graphics} 

\begin{document}


\title{\bf Termination and confluence\\ of higher-order rewrite systems
\vspace*{2mm}}

\author{Fr\'ed\'eric Blanqui\vspace*{1mm}\\
{\small LRI, Universit\'e de Paris-Sud}\\
{\small B\^at. 490, 91405 Orsay, France}\\
{\small tel: (33) 1.69.15.42.35 ~ fax: (33) 1.69.15.65.86}\\
{\small\tt Frederic.Blanqui@lri.fr}\\
{\small\tt http://www.lri.fr/\home{}blanqui/}
}

\date{}

\maketitle

\noindent {\bf Abstract:}
In the last twenty years, several approaches to higher-order rewriting
have been proposed, among which Klop's Combinatory Rewrite Systems
(CRSs), Nipkow's Higher-order Rewrite Systems (HRSs) and Jouannaud and
Okada's higher-order algebraic specification languages, of which only
the last one considers typed terms. The later approach has been
extended by Jouannaud, Okada and the present author into Inductive
Data Type Systems (IDTSs). In this paper, we extend IDTSs with the CRS
higher-order pattern-matching mechanism, resulting in simply-typed
CRSs. Then, we show how the termination criterion developed for IDTSs
with first-order pattern-matching, called the General Schema, can be
extended so as to prove the strong normalization of IDTSs with
higher-order pattern-matching. Next, we compare the unified approach
with HRSs. We first prove that the extended General Schema can also
be applied to HRSs. Second, we show how Nipkow's higher-order critical
pair analysis technique for proving local confluence can be applied to
IDTSs.

\comment{
~

Appendices \ref{app-beta-idts}, \ref{app-gs-hrs} and \ref{app-proofs}
(proofs) are available from the web page.
}



\section{Introduction}

In 1980, after a work by Aczel \cite{aczel78tr}, Klop introduced the
Combinatory Rewrite Systems (CRSs) \cite{klop80thesis, klop93tcs}, to
generalize both first-order term rewriting and rewrite systems with
bound variables like Church's $\l$-calculus.

In 1991, after Miller's decidability result of the pattern unification
problem \cite{miller89elp}, Nipkow introduced Higher-order Rewrite
Systems (HRSs) \cite{nipkow91lics} (called Pattern Rewrite Systems
(PRSs) in \cite{mayr98tcs}), to investigate the metatheory of logic
programming languages and theorem provers like $\l$Prolog
\cite{miller88lp} or Isabelle \cite{paulson94lncs}. In particular, he
extended to the higher-order case the decidability result of Knuth and
Bendix about local confluence of first-order term rewrite systems.

At the same time, after the works of Breazu-Tannen
\cite{breazu88lics}, Breazu-Tannen and Gallier \cite{breazu89icalp}
and Okada \cite{okada89issac} on the combination of Church's
simply-typed $\l$-calculus with first-order term rewriting, Jouannaud
and Okada introduced higher-order algebraic specification languages
\cite{jouannaud91lics, jouannaud97tcs} to provide a computational
model for typed functional languages extended with first-order and
higher-order rewrite definitions. Later, together with the present
author, they extended these languages with (strictly positive)
inductive types, leading to Inductive Data Type Systems (IDTSs)
\cite{blanqui98tcssub}. This approach has also been adapted to richer
type disciplines like Coquand and Huet's Calculus of Constructions
\cite{barbanera97jfp, blanqui99rta}, in order to extend the equality
used in proof assistants based on the Curry-De~Bruijn-Howard
isomorphism like Coq \cite{coq} or Lego \cite{lego}.

~

Although CRSs and HRSs seem quite different, they have been precisely
compared by van Oostrom and van Raamsdonk \cite{oostrom93hoa}, and
shown to have the same expressive power, CRSs using a more lazy
evaluation strategy than HRSs. On the other hand, although IDTSs seem
very close in spirit to CRSs, the relation between both systems has not
been clearly stated yet.

Other approaches have been proposed like Wolfram's Higher-Order Term
Rewriting Systems (HOTRSs) \cite{wolfram90thesis}, Khasidashvili's
Expression Reduction Systems (ERSs) \cite{khasidashvili90},
Takahashi's Conditional Lambda-Calculus (CLC) \cite{takahashi93tlca},
\ldots (see \cite{oostrom94thesis}). To tame this proliferation, van
Oostrom and van Raamsdonk introduced Higher-Order Rewriting Systems
(HORSs) \cite{oostrom94thesis, raamsdonk96thesis} in which the
matching procedure is a parameter called ``substitution calculus''. It
appears that most of the known approaches can be obtained by using an
appropriate substitution calculus. Van Oostrom proved important
confluence results for HORSs whose substitution calculus fulfill some
conditions, hence factorizing the existing proofs for the different
approaches.

~

Many results have been obtained so far about the confluence of CRSs
and HRSs. On the other hand, for IDTSs, termination was the target of
research efforts. A powerful and decidable termination criterion has
been developed by Jouannaud, Okada and the present author, called the
General Schema \cite{blanqui98tcssub}.

So, one may wonder whether the General Schema may be applied to HRSs,
and whether Nipkow's higher-order critical pair analysis technique for
proving local confluence of HRSs may be applied to IDTSs.

~

This paper answers positively both questions. However, we do not
consider the {\em critical interpretation} introduced in
\cite{blanqui98tcssub} for dealing with function definitions over
strictly positive inductive types (like Brouwer's ordinals or process
algebra). In Section~\ref{sec-newdef}, we show how IDTSs relate to
CRSs and extend IDTSs with the CRS higher-order pattern-matching
mechanism, resulting in simply-typed CRSs. In
Section~\ref{sec-schema}, we adapt the General Schema to this new
calculus and prove in Section~\ref{sec-termination} that the rewrite
systems that follow this schema are strongly normalizing (every
reduction sequence is finite).  In Section~\ref{sec-hrs}, we show that
it can be applied to HRSs. In Section~\ref{sec-confluence}, we show
that Nipkow's higher-order critical pair analysis technique can be
applied to IDTSs.

~

For proving the termination of a HRS, other criteria are available.
Van de Pol extended to the higher-order case the use of strictly
monotone interpretations \cite{vandepol95tlca}. This approach is of
course very powerful but it cannot be automated. In
\cite{jouannaud99lics}, Jouannaud and Rubio defined an extension to
the higher-order case of Dershowitz' Recursive Path Ordering (HORPO)
exploiting the notion of computable closure introduced in
\cite{blanqui98tcssub} by Jouannaud, Okada and the present author for
defining the General Schema. Roughly speaking, the General Schema may
be seen as a non-recursive version of HORPO. However, HORPO has not
yet been adapted to higher-order pattern-matching.



\section{Preliminaries}
\label{sec-prelim}

We assume that the reader is familiar with simply-typed $\l$-calculus
\cite{barendregt93book}. The set $T(\cB)$ of {\em types} $s,t, \ldots$
generated from a set $\cB$ of {\em base types} ${\tt s}, {\tt t},
\ldots$ (in bold font) is the smallest set built from $\cB$ and the
function type constructor $\a$. We denote by $FV(u)$ the set of free
variables of a term $u$, $u \!\ad_\b$ (resp. $u \!\au^\e$) the
$\b$-normal form of $u$ (resp. the $\e$-long form of $u$).

We use a postfix notation for the application of substitutions, $\{
x_1 \to u_1, \ldots,\\ x_n \to u_n \}$ for denoting the substitution
$\t$ such that $x_i\t = u_i$ for each $i \in \{ 1, \ldots, n \}$, and
$\t \uplus \{ x \to u \}$ when $x \notin dom(\t)$, for denoting the
substitution $\t'$ such that $x\t' = u$ and $y\t' = y\t$ if $y \neq
x$. The domain of a substitution $\t$ is the set $dom(\t)$ of
variables $x$ such that $x\t \neq x$. Its codomain is the set $cod(\t)
= \{ x\t ~|~ x \in dom(\t) \}$.

Whenever we consider abstraction operators, like $\l\_.\_$ in
$\l$-calculus, we work modulo $\alpha$-conversion, \ie modulo renaming
of bound variables. Hence, we can always assume that, in a term, the
bound variables are pairwise distinct and distinct from the free
variables. In addition, to avoid variable capture when applying a
substitution $\t$ to a term $u$, we can assume that the free variables
of the terms of the codomain of $\t$ are distinct from the bound
variables of $u$.

We use words over positive numbers for denoting positions in a term.
With a symbol $f$ of fixed arity, say $n$, the positions of the
arguments of $f$ are the numbers $i \in \{ 1, \ldots, n \}$. We will
denote by $Pos(u)$ the set of positions in a term $u$. The subterm at
position $p$ is denoted by $u|_p$. Its replacement by another term $v$
is denoted by $u[v]_p$.

For the sake of simplicity, we will often use vector notations for
denoting comma- or space-separated sequences of objects. For example,
$\{ \vec{x} \to \vec{u} \}$ will denote $\{ x_1 \to u_1, \ldots, x_n
\to u_n \}$, $n = |\vec{u}|$ being the length of $\vec{u}$. Moreover,
some functions will be naturally extended to sequences of objects. For
example, $FV(\vec{u})$ will denote $\bigcup_{1\le i\le n} FV(u_i)$ and
$\vec{u}\t$ the sequence $u_1\t \ldots u_n\t$.



\newcommand{\la}{\l^\a}
\newcommand{\ul}{\u{\l}}
\newcommand{\M}{M\!\L}
\newcommand{\var}{V\!ar}

\section{Extending IDTSs with higher-order pattern-matching {\em \`a la} CRS}
\label{sec-newdef}


In a Combinatory Rewrite System (CRS) \cite{klop93tcs}, the terms are
built from variables $x, y, \ldots$ function symbols $f, g, \ldots$ of
fixed arity and an abstraction operator $[\_]\_$ such that, in $[x]u$,
the variable $x$ is bound in $u$. On the other hand, left-hand and
right-hand sides of rules are not only built from variables, function
symbols and the abstraction operator like terms, but also from
metavariables $Z, Z', \ldots$ of fixed arity. In the left-hand sides
of rules, the metavariables must be applied to distinct bound
variables (a condition similar to the one for patterns {\em \`a la}
Miller \cite{mayr98tcs}). By convention, a term $Z(x_{i_1}, \ldots,
x_{i_k})$ headed by $[x_1], \ldots, [x_n]$ can be replaced only by a
term $u$ such that $FV(u) \cap \{ x_1, \ldots, x_n \} \sle \{ x_{i_1},
\ldots, x_{i_k} \}$.

For example, in a left-hand side of the form $f([x][y]Z(x))$, the
metaterm $Z(x)$ stands for a term in which $y$ cannot occur free, that
is, the metaterm $[x][y]Z(x)$ stands for a function of two variables
$x$ and $y$ not depending on $y$.

The $\l$-calculus itself may be seen as a CRS with the symbol $@$ of
arity 2 for the application, the CRS abstraction operator $[\_]\_$
standing for $\l$, and the rule

\begin{center}
$@([x]Z(x),Z') \a Z(Z')$
\end{center}

\noindent
for the $\b$-rewrite relation. Indeed, by definition of the CRS
substitution mechanism, if $Z(x)$ stands for some term $u$ and $Z'$
for some other term $v$, then $Z(Z')$ stands for $u \{ x \to v \}$.

In \cite{blanqui98tcssub}, Inductive Data Type Systems (IDTSs) are
defined as extensions of the simply-typed $\l$-calculus with function
symbols of fixed arity defined by rewrite rules. So, an IDTS may be
seen as the sub-CRS of well-typed terms, in which the free variables
occuring in rewrite rules are metavariables of arity $0$, and only
$\b$ really uses the CRS substitution mechanism.

As a consequence, restricting matching to first-order matching clearly
leads to non-confluence. For example, the rule

\begin{center}
$D(\l x. sin(F ~x)) \a \l x. (D(F) ~x) \!\times\! cos(F ~x)$
\end{center}

\noindent
defining a formal differential operator $D$ over a function of the
form $sin \circ F$, cannot rewrite a term of the form $D(\l x.
sin(x))$ since $x$ is not of the form $(u ~x)$.

On the other hand, in the CRS approach, thanks to the notions of
metavariable and substitution, $D$ may be properly defined with the rule

\begin{center}
$D([x]sin(F(x))) \a [x] \, @(D([y]F(y)),x) \!\times\! cos(F(x))$
\end{center}

\noindent
where $F$ is a metavariable of arity 1.

This leads us to extend IDTSs with the CRS notions of metavariable and
substitution, hence resulting in simply-typed CRSs.


\begin{dfn}[IDTS - new definition]
An {\em IDTS-alphabet} $\cA$ is a 4-tuple\\ $(\cB, \cX, \cF, \cZ)$ where:
\begin{lst}{--}
\item $\cB$ is a set of {\em base types},
\item $\cX$ is a family $(X_t)_{t \in T(\cB)}$ of sets of {\em
    variables},
\item $\cF$ is a family $(F_{s_1, \ldots, s_n, s})_{n \ge 0, s_1,
    \ldots, s_n, s \in T(\cB)}$ of sets of {\em function symbols},
\item $\cZ$ is a family $(Z_{s_1, \ldots, s_n, s})_{n \ge 0, s_1,
    \ldots, s_n, s \in T(\cB)}$ of sets of {\em metavariables},
\end{lst}

\noindent
such that all the sets are pairwise disjoint.

The set of {\em IDTS-metaterms} over $\cA$ is $\cI(\cA) = \bigcup_{t
  \in T(\cB)} \cI_t$ where $\cI_t$ are the smallest sets such that:
\begin{enum}
\item $X_t \sle \cI_t$,
\item if $x \in X_s$ and $u \in \cI_t$, then $[x] u \in \cI_{s \a t}$,
\item if $f \in F_{s_1, \ldots, s_n, s}$, $u_1 \in \cI_{s_1}, \ldots,
  u_n \in \cI_{s_n}$, then $f(u_1, \ldots, u_n) \in \cI_s$.
\item if $Z \in Z_{s_1, \ldots, s_n, s}$, $u_1 \in \cI_{s_1}, \ldots,
  u_n \in \cI_{s_n}$, then $Z(u_1, \ldots, u_n) \in \cI_s$.
\end{enum}

\noindent
We say that a metaterm $u$ is {\em of type $t \in T(\cB)$} if $u \in
\cI_t$. The set of metavariables occuring in a metaterm $u$ is denoted
by $\var(u)$. A {\em term} is a metaterm with no metavariable.

A metaterm $l$ is an {\em IDTS-pattern} if every metavariable occuring
in $l$ is applied to a sequence of distinct bound variables.

An {\em IDTS-rewrite rule} is a pair $l \a r$ of metaterms such that:
\begin{enum}
\item $l$ is an IDTS-pattern,
\item $l$ is headed by a function symbol,
\item $\var(r) \sle \var(l)$,
\item $r$ has the same type as $l$,
\item $l$ and $r$ are closed ~($FV(l) = FV(r) = \emptyset$).
\end{enum}

\noindent
An {$n$-ary substitute of type $s_1 \a \ldots \a s_n \a s$} is an
expression of the form $\ul(\vec{x}). u$ where $\vec{x}$ are distinct
variables of respective types $s_1, \ldots, s_n$ and $u$ is a term of
type $s$. An {\em IDTS-valuation} $\s$ is a type-preserving map
associating an $n$-ary substitute to each metavariable of arity $n$.
Its (postfix) application to a metaterm returns a term defined as
follows:
\begin{lst}{--}
\item $x\s = x$
\item $([x] u)\s = [x] u\s \quad (x \notin FV(cod(\s)))$
\item $f(\vec{u})\s = f(\vec{u}\s)$
\item $Z(\vec{u})\s = v \{ \vec{x} \to \vec{u}\s \}$ ~if ~$\s(Z) =
  \ul(\vec{x}). v$
\end{lst}

\noindent
An {\em IDTS} $\cI$ is a pair $(\cA, \cR)$ where $\cA$ is an
IDTS-alphabet and $\cR$ is a set of IDTS-rewrite rules over $\cA$.
Its corresponding rewrite relation $\a_\cI$ is the subterm compatible
closure of the relation containing every pair $l\s \a r\s$ such that
$l \a r \in \cR$ and $\s$ is an IDTS-valuation over $\cA$.
\end{dfn}


The following class of IDTSs will interest us especially:

\begin{dfn}[$\b$-IDTS]
  An IDTS $(\cA, \cR)$ where $\cA = (\cB, \cX, \cF, \cZ)$ is a {\em
    $\b$-IDTS} if, for every pair $s,t \in T(\cB)$, there is:
\begin{enum}
\item a function symbol $@_{s,t} \in F_{s\a t, s, t}$,
\item a rule $\b_{s,t} = @([x]Z(x), Z') \a Z(Z') \in \cR$,
\end{enum}

\noindent
and no other rule has a left-hand side headed by $@$.

Given an IDTS $\cI$, we can always add new symbols and new rules so as
to obtain a $\b$-IDTS. We will denote by $\b\cI$ this {\em
  $\b$-extension} of $\cI$.
\end{dfn}

For short, we will denote $@( \ldots @(@(v, u_1), u_2), \ldots, u_n)$
by $@(v, \vec{u})$.

The strong normalization of $\b\cI$ trivially implies the strong
normalization of $\cI$. However, the study of $\b\cI$ seems a
necessary step because the application symbol $@$ together with the
rule $\b$ are the essence of the substitution mechanism. Should we
replace in the right-hand sides of the rules every metaterm of the
form $Z(\vec{u})$ by $@([\vec{x}]Z(\vec{x}), \vec{u})$, the system
would lead to the same normal forms.

In Appendix~\ref{app-beta-idts}, we list some results about the
relations between $\cI$ and $\b\cI$.



\section{Definition of the General Schema}
\label{sec-schema}

All along this section and the following one, we fix a given $\b$-IDTS
$\cI = (\cA, \cR)$.  Firstly, we adapt the definition of the General
Schema given in \cite{blanqui98tcssub} to take into account the notion
of metavariable. Then, we prove that if the rules of $\cR$ follow this
schema, then $\a_\cI$ is strongly normalizing.

The General Schema is a syntactic criterion which ensures the strong
normalization of IDTSs. It has been designed so as to allow a strong
normalization proof by the technique of {\em computability predicates}
introduced by Tait for proving the normalization of the simply-typed
$\l$-calculus \cite{tait67jsl,hindley86book}. Hereafter, we only give
basic definitions. The reader will find more details in
\cite{blanqui98tcssub}.

Given a rule with left-hand side $f(\vec{l})$, we inductively define a
set of admissible right-hand sides that we call the {\em computable
  closure} of $\vec{l}$, starting from the {\em accessible}
metavariables of $\vec{l}$. The main problem will be to prove that the
computable closure is indeed a set of ``computable'' terms whenever
the terms in $\vec{l}$ are ``computable''. This is the objective of
Lemma~\ref{lem-comp-clos-correct} below. The notion of computable
closure has been first introduced by Jouannaud, Okada and the present
author in \cite{blanqui98tcssub,blanqui99rta} for defining the General
Schema, but it has been also used by Jouannaud and Rubio in
\cite{jouannaud99lics} for strengthening their Higher-Order Recursive
Path Ordering.


~

For each base type ${\tt s}$, we assume given a set $C_{\tt s} \sle
\bigcup_{p \ge 0, s_1, \ldots, s_p \in T(\cB)} F_{s_1, \ldots, s_p,
  {\tt s}}$ whose elements are called the {\em constructors} of ${\tt
  s}$. When a function symbol is a constructor, we may denote it by
the lower case letters $c, d, \ldots$

This induces the following relation on base types: ${\tt t}$ {\em
  depends on} ${\tt s}$ if there is a constructor $c \in C_{\tt t}$
such that ${\tt s}$ occurs in the type of one of the arguments of $c$.
Its reflexive and transitive closure $\le_\cB$ is a quasi-ordering
whose associated equivalence relation (resp. strict ordering) will
be denoted by $=_\cB$ (resp. $<_\cB$).

We say that a constructor $c \in C_{\tt s}$ is {\em positive} if every
base type ${\tt t} =_\cB {\tt s}$ occurs only at positive positions
(wrt. the type constructor $\a$) into the types of the arguments of
$c$. $c$ is {\em basic} if it is positive and has no functional
arguments. A type is {\em positive} (resp. {\em basic}) if all its
constructors are positive (resp. basic).


\begin{dfn}[Accessible subterms]
\label{dfn-acc}
The set $Acc(v)$ of {\em accessible subterms} of a metaterm $v$ is the
smallest set such that:

\begin{enum}
\item $v \in Acc(v)$
\item if $[x] u \in Acc(v)$ then $u \in Acc(v)$
\item if $c(\vec{u}) \in Acc(v)$ then each $u_i \in Acc(v)$
\item if $f(\vec{u}) \in Acc(v)$ and $u_i$ is of basic type then $u_i
  \in Acc(v)$
\item if $@(u,x) \in Acc(v)$, $x \notin FV(u) \cup FV(v)$ then $u \in
  Acc(v)$
\item if $@(x,\vec{u}) \in Acc(v)$, $x \notin FV(\vec{u}) \cup
  FV(v)$ then each $u_i \in Acc(v)$.
\end{enum}

\noindent
By abuse of notation, we will say that a metavariable $Z$ is {\em
  accessible} in $v$ if there are distinct bound variables $\vec{x}$
such that $Z(\vec{x}) \in Acc(v)$.
\end{dfn}

For example, $F$ is accessible in $v = [x] sin(F(x))$ since
$sin(F(x))$ is accessible in $v$ by (2), and thus, $F(x)$ is
accessible in $v$ by (3).

Compared to \cite{blanqui98tcssub}, we express the accessibility with
respect to a fixed $v$. This has no consequence on the definition of
computable closure since, among the accessible subterms, only the free
variables (here, the metavariables) are taken into account.
Accessibility enjoys the following property:


\begin{prop}
\label{prop-acc}
If $u \in Acc(v)$ then $u\s \in Acc(v\s)$.
\end{prop}


For proving termination, we are led to compare the arguments of a
function symbol with the arguments of the recursive calls generated by
its reductions. To this end, each function symbol $f \in \cF$ is
equipped with a {\em status} $stat_f$ which specifies how to make the
comparison as a simple combination of multiset and lexicographic
comparisons. Then, an ordering on terms $\le$ is easily extended to an
ordering on sequences of terms $\le\stat$. The reader will find
precise definitions in \cite{blanqui98tcssub}. To fix an idea, one can
assume that $\le\stat$ is the lexicographic extension $\le\lex$ or the
multiset extension $\le\mul$ of $\le$. We will denote by $\le\statt$
(resp.  $\le\state$) the strict ordering (resp. equivalence relation)
associated to $\le\stat$. $\le\statt$ is well-founded if the strict
ordering associated to $\le$ is well-founded.

$\cR$ induces the following relation on function symbols: $g$ {\em
  depends on} $f$ if there is a rewrite rule defining $g$ (\ie whose
left-hand side is headed by $g$) in the right-hand side of which $f$
occurs. Its reflexive and transitive closure is a quasi-ordering
denoted by $\le_\cF$ whose associated equivalence relation (resp.
strict ordering) will be denoted by $=_\cF$ (resp. $<_\cF$).


Finally, we will do the following

~

\noindent
{\bf Assumptions (A)}
\begin{lst}{}
\item [(1)] every constructor is positive
\item [(2)] no left-hand side of rule is headed by a constructor
\item [(3)] both $>_\cB$ and $>_\cF$ are well-founded
\item [(4)] $stat_f = stat_g$ whenever $f =_\cF g$
\end{lst}{}

~

The first assumption comes from the fact that, from non-positive
inductive types, it is possible to build non-terminating terms
\cite{mendler87thesis}. The second assumption ensures that if a
constructor-headed term is computable, then its arguments are
computable too. The third assumption ensures that types and function
definitions are not cyclic. The fourth assumption says that the
arguments of equivalent symbols must be compared in the same way.


For comparing the arguments, the subterm ordering $\tle$ used in
\cite{blanqui98tcssub} is not satisfactory anymore because of the
metavariables which must be applied to some arguments. For example,
$[x] F(x)$ is not a subterm of $[x] sin(F(x))$. This can be repaired
by using the following ordering.

\newcommand{\hle}{\h{\tle}}
\newcommand{\hlt}{\h{\tlt}}
\newcommand{\hge}{\h{\tge}}
\newcommand{\hgt}{\h{\tgt}}

\begin{dfn}[Covered-subterm ordering]
  We say that a metaterm $u$ is a {\em covered-subterm} of a metaterm
  $v$, written $u ~\hle~ v$, if there are two positions $p \in Pos(v)$
  and $q \in Pos(v|_p)$ such that (see the figure):
\begin{lst}{--}
\item $u = v[v|_{pq}]_p$,
\item $\all r < p$, $v|_r$ is headed by an abstraction,
\item $\all r < q$, $v|_{pr}$ is headed by a function symbol (which
  can be a constructor).
\end{lst}
\end{dfn}

\begin{figure}[ht]
\begin{center}
\scalebox{.3}{\includegraphics{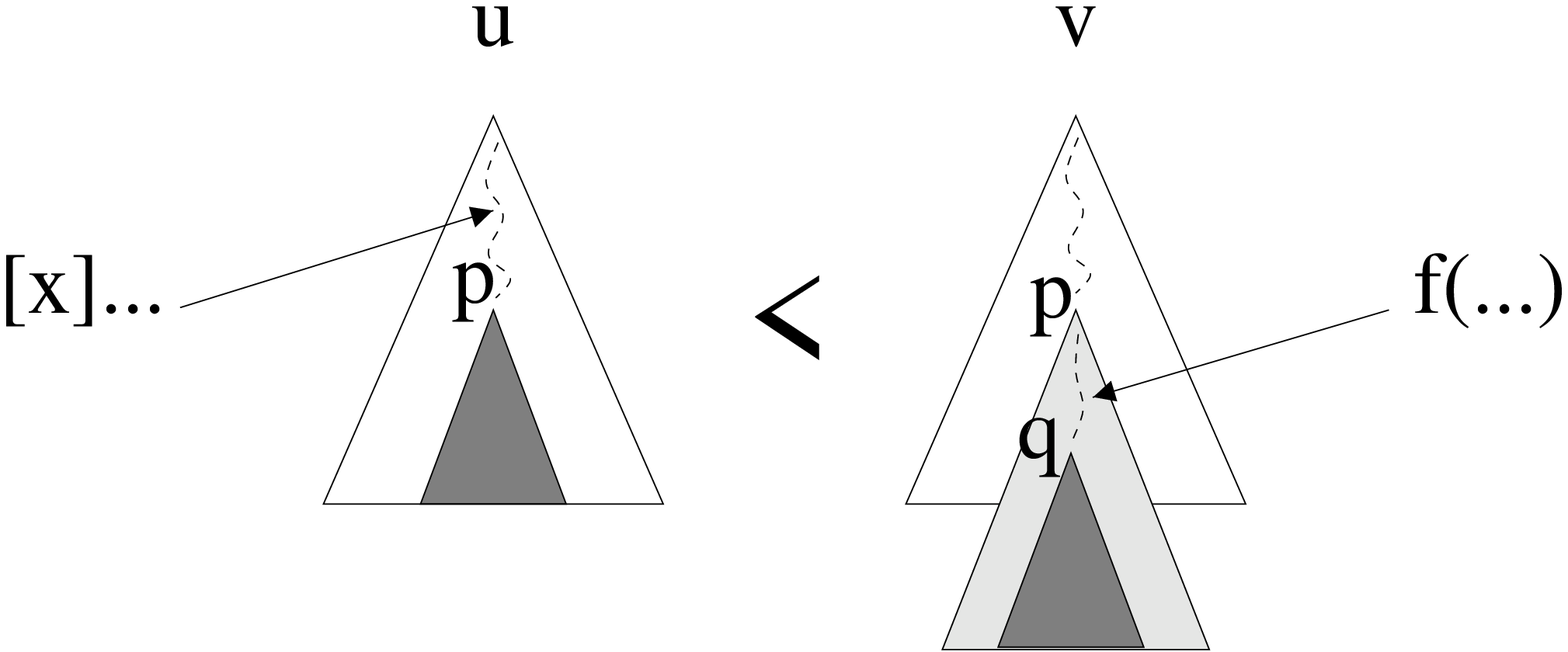}}
\end{center}
\end{figure}



\begin{prop}
\label{prop-cov-ordering}
\hfill
\begin{enum}
\item $\hgt$ is stable by valuation: if $u ~\hgt~ v$ and $\s$ is a
  valuation, then $u\s ~\hgt~ v\s$.
\item $\hgt$ is stable by substitution: if $u ~\hgt~ v$ and $\t$ is a
  substitution, then $u\t ~\hgt~ v\t$.
\item $\hgt$ commutes with $\a$: if $u ~\hgt~ v$ and $v \a w$ then
  there is a term $v'$ such that $u \a v'$ and $v' ~\hgt~ w$.
\end{enum}
\end{prop}


Finally, we come to the definition of computable closure.

\newcommand{\C}{{\cal CC}}

\begin{dfn}[Computable closure]
\label{dfn-comp-closure}
Given a function symbol $f \in F_{s_1, \ldots, s_n, s}$, the {\em
  computable closure} $\C_f(\vec{l})$ of a metaterm $f(\vec{l})$ is
the least set $\C$ such that:

\begin{enum}
\item if $Z \in Z_{t_1, \ldots, t_p, t}$ is accessible in $\vec{l}$
  and $\vec{u}$ are $p$ metaterms of $\C$ of respective types $t_1,
  \ldots, t_p$, then $Z(\vec{u}) \in \C$;
\item if $x \in X_t$ then $x \in \C$;
\item if $c \in C_{\tt t} \cap F_{t_1, \ldots, t_p, {\tt t}}$ and
  $\vec{u}$ are $p$ metaterms of $\C$ of respective types $t_1,
  \ldots, t_p$, then $c(\vec{u}) \in \C$;
\item if $u$ and $v$ are two metaterms of $\C$ of respective types
  $s\a t$ and $s$ then $@(u,v) \in \C$;
\item if $u\in \C$ then $[x]u \in \C$;
\item if $h \in F_{t_1, \ldots, t_p, t}$, $h <_\cF f$ and $\vec{w}$
  are $p$ metaterms of $\C$ of respective types $t_1, \ldots, t_p$,
  then $h(\vec{w}) \in \C$;
\item if $g \in F_{t_1, \ldots, t_p, t}$, $g =_\cF f$ and $\vec{u}$
  are $p \ge 1$ metaterms of $\C$ of respective types $t_1, \ldots,
  t_p$ such that $\vec{u} ~~\hle\statt~ \vec{l}$, then $g(\vec{u})
  \in \C$.
\end{enum}
\end{dfn}

Note that we do not consider in case (7) the notion of {\em critical
  interpretation} introduced in \cite{blanqui98tcssub} for proving the
termination of function definitions over strictly positive types (like
Brouwer's ordinals or process algebra).


\begin{dfn}[General Schema]
  A rewrite rule $f(\vec{l}) \a r$ follows the {\em General Schema} GS
  if $r \in \C_f(\vec{l})$.
\end{dfn}

A first example is given by the rule $\b$ itself: $@([x]Z(x),Z') \a
Z(Z')$ ($Z$ and $Z'$ are both accessible).

$D([x] sin(F(x))) \a [x] @(D([y] F(y)), x) \!\times\!  cos(F(x))$
also follows the General Schema since $x$ and $y$ belong to the
computable closure of $[x] sin(F(x))$ by (2), hence $F(x)$ and $F(y)$
by (1) since $F$ is accessible in $[x] sin(F(x))$, $[y] F(y)$ by (5),
$D([y] F(y))$ by (7) since $[y] F(y)$ is a strict covered-subterm of
$[x] sin(F(x))$, $@(D([y] F(y)), x)$ by (4), $cos(F(x))$ by (3),
$@(D([y] F(y)), x) \!\times\! cos(F(x))$ by (6) and the whole
right-hand side by (5).



\section{Termination proof}
\label{sec-termination}

The termination proof follows Tait's technique of computability
predicates \cite{tait67jsl, hindley86book}. Computability predicates
are sets of strongly normalizable terms satisfying appropriate
conditions. For each type, we define an interpretation which is a
computability predicate and we prove that every term is computable,
\ie it belongs to the interpretation of its type. For precise
definitions, see \cite{blanqui98tcssub}.


The main things to know are:
\begin{lst}{--}
\item Computability implies strong normalizability.
\item If $u$ is a term of type $s\a t$, then it is computable iff, for
  every computable term $v$ of type $s$, $@(u,v)$ is computable.
\item Computability is preserved by reduction.
\item A term is {\em neutral} if it is neither constructor-headed nor
  an abstraction. A neutral term $u$ is computable if all its
  immediate reducts are computable.
\item A constructor-headed term $c(\vec{u})$ is computable iff all the
  terms in $\vec{u}$ are computable.
\item For basic types, computability is equivalent to strong
  normalizability.
\end{lst}


\begin{dfn}[Computable valuation]
  A substitution is {\em computable} if all the terms of its codomain
  are computable. A substitute $\ul(\vec{x}). u$ is {\em computable}
  if, for any computable substitution $\t$ such that $dom(\t) \sle \{
  \vec{x} \}$, $u\t$ is computable. Finally, a valuation $\s$ is {\em
    computable} if, for every metavariable $Z$, the substitute $\s(Z)$
  is computable.
\end{dfn}


\begin{lem}[Compatibility of accessibility with computability]
\label{lem-compat-acc-comp}
If $u \in\\ Acc(v)$ and $v$ is computable, then for any computable
substitution $\t$ such that $dom(\t) \cap FV(v) = \emptyset$, $u\t$ is
computable.
\end{lem}

\begin{prf}
  By induction on $Acc(v)$. Without loss of generality, we can assume
  that $dom(\t) \sle FV(u)$ since $u\t = u\t|_{FV(u)}$.
\begin{prfenum}
\item Immediate.
\item $\t$ is of the form $\t' \uplus \{ x \to x\t \}$ where $dom(\t')
  \cap FV(v) = \emptyset$. By induction hypothesis, $([x]u)\t'$ is
  computable. By taking $x$ away from $FV(cod(\t'))$, $([x]u)\t' =
  [x]u\t'$ and $u\t = u\t' \{ x \to x\t \}$ is a reduct of
  $@([x]u\t',x\t)$, hence it is computable since $x\t$ is computable.
\item By induction hypothesis, $c(\vec{u})\t = c(\vec{u}\t)$ is
  computable. Hence, by definition of the interpretation for inductive
  types, $u_i\t$ is computable.
\item By induction hypothesis, $f(\vec{u})\t = f(\vec{u}\t)$ is
  computable. Hence $u_i\t$ is strongly normalizable, and since, for
  terms of basic type, computability is equivalent to strong
  normalizability, $u_i\t$ is computable.
\item $u$ must be of type $s \a t$. So, let $w$ be a computable term
  of type $s$. Since $x \notin FV(u)$, $x \notin dom(\t)$. Then, let
  $\t' = \t \uplus \{ x \to w \}$. $\t'$ is computable and $dom(\t')
  \cap FV(v) = \emptyset$ since $x \notin FV(v)$. Hence, by induction
  hypothesis, $@(u, x)\t' = @(u\t, w)$ is computable.
\item Since $x \notin FV(u)$, $x \notin dom(\t)$. Then, let $\t' = \t
  \uplus \{ x \to [\vec{y}]y_i \}$, $[\vec{y}]y_i$ being the $i$-th
  projection. $\t'$ is computable and $dom(\t') \cap FV(v) =
  \emptyset$ since $x \notin FV(v)$.  Hence, by induction hypothesis,
  $@(x,\vec{u})\t' = @([\vec{y}]y_i,\vec{u}\t)$ is computable and its
  $\b$-reduct $u_i\t$ also.
\end{prfenum}
\end{prf}


\begin{cor}
\label{cor-compat-acc-comp}
Let $l$ be a pattern, $v$ a term and $\s$ a valuation such that $l\s =
v$. If $Z$ is accessible in $l$ and $v$ is computable, then $\s(Z)$ is
computable.
\end{cor}


For proving Lemma~\ref{lem-comp-fun-symb} below, we will reason by
induction on $(f, \vec{u})$ with the ordering $\cge \,= (\ge_\cF,
\a\mul \cup ~\hge\statt)\lex$, $\vec{u}$ being strongly normalizable
arguments of $f$. Since $\hgt$ commutes with $\a$, we can prove that
$\hge\statt\!\!\a\mul$ is included into $\a\mul\!\!^{0,1} ~\hge\statt$
where $\a\mul\!\!^{0,1}$ means zero or one $\a\mul$-step. This implies
that $\a\mul \cup ~\hge\statt$ is well-founded since:

\begin{lem}
\label{lem-commut}
If $a$ and $b$ are two well-founded relations such that $ab \sle b^*a$
then $a \cup b$ is well-founded.
\end{lem}

Therefore the strict ordering $\cgt$ associated to $\cge$ is
well-founded since $>_\cF$ is assumed to be well-founded. Now, we can
prove the correctness of the computable closure.

\begin{lem}[Computable closure correctness]
\label{lem-comp-clos-correct}
Let $f(\vec{l})$ be a pattern. Assume that $\s$ is a computable
valuation and that the terms in $\vec{l}\s$ are computable. Assume
also that, for every function symbol $h$ and sequence of computable
terms $\vec{w}$ such that $(f, \vec{l}\s) \cgt (h,\vec{w})$,
$h(\vec{w})$ is computable. Then, for every $r \in \C_f(\vec{l})$,
$r\s$ is computable.
\end{lem}

\begin{prf}
  The proof, by induction on $\C_f(\vec{l})$, is quite similar to the
  one given in \cite{blanqui98tcssub} except that, now, one has to
  deal with valuations instead of substitutions. The main difference
  is in case (1) for metavariables. We only give this case.  A full
  proof can be found in Appendix~\ref{app-proofs}.

In fact, we prove that, for any computable valuation $\s$ such that
$FV(cod(\s)) \cap FV(r) = \emptyset$, for any computable substitution
$\t$ such that $dom(\t) \sle FV(r)$ and for any $r \in \C_f(\vec{l})$,
$r\s\t = r\t\s$ is computable.

\begin{prfenum}
\item $r = Z(\vec{v})$ where $Z$ is a metavariable accessible in
  $\vec{l}$ and $\vec{v}$ are metaterms of $\C$. We first prove it for
  a special case and then for the general case.
  \begin{lstx}{}{\usecounter{enumii}%
     \renewcommand{\makelabel}{(\alph{enumii})}%
     \labelsep=-3mm\itemindent=3mm}
 \item $\vec{v}$ is a sequence of distinct bound variables, say
   $\vec{x}$. Without loss of generality, we can assume that $\s(Z) =
   \ul(\vec{x}). w$. Then, $r\s\t = w\t$. Since $\s$ is computable and
   $dom(\t) \sle \{\vec{x}\} = FV(r)$, $w\t$ is computable.
 \item $r\s\t$ is a $\b$-reduct of the term $@([\vec{x}]
   Z(\vec{x})\s\t, \vec{v}\s\t)$ where $\vec{x}$ are fresh distinct
   variables. By case (1a) and (5), $[\vec{x}]Z(\vec{x})\s\t$ is
   computable and since, by induction hypothesis, the terms in
   $\vec{v}\s\t$ are also computable, $r\s\t$ is computable.
  \end{lstx}
\end{prfenum}
\end{prf}

\begin{lem}[Computability of function symbols]
\label{lem-comp-fun-symb}
  If all the rules satisfy the General Schema then, for every
  function symbol $f$, $f(\vec{u})$ is computable whenever the terms
  in $\vec{u}$ are computable.
\end{lem}

\begin{prf}
  If $f$ is a constructor then this is immediate since the terms in
  $\vec{u}$ are computable by assumption. Assume now that $f$ is a
  function symbol.  Since $f(\vec{u})$ is neutral, to prove that
  $f(\vec{u})$ is computable, it suffices to prove that all its
  immediate reducts are computable. We prove this by induction on $(f,
  \vec{u})$ with $\cgt$ as well-founded ordering.

Let $v$ be an immediate reduct of $f(\vec{u})$. $v$ is either a
head-reduct of $f(\vec{u})$ or of the form $f(u_1, \ldots, u_i',
\ldots, u_n)$ with $u_i'$ being an immediate reduct of $u_i$.

In the latter case, as computability predicates are stable by
reduction, $u_i'$ is computable. Hence, since $(f, u_1 \ldots u_i'
\ldots u_n) \clt (f, \vec{u})$, by induction hypothesis, $f(u_1,
\ldots, u_i', \ldots, u_n)$ is computable.
  
In the former case, there is a rule $f(\vec{l}) \a r$ and a valuation
$\s$ such that $\vec{u} = \vec{l}\s$ and $v = r\s$. By definition of
the computable closure, and since $\var(r) \sle \var(\vec{l})$, every
metavariable occuring in $r$ is accessible in $\vec{l}$. Hence, since
the terms in $\vec{l}\s$ are computable, by
Corollary~\ref{cor-compat-acc-comp}, $\s|_{\var(r)}$ is computable.
Therefore, by Lemma~\ref{lem-comp-clos-correct}, $r\s =
r\s|_{\var(r)}$ is computable.
\end{prf}


\begin{thm}[Strong normalization]
\label{thm-sn}
  Let $\cI = (\cA, \cR)$ be a $\b$-IDTS satisfying the assumptions
  (A). If all the rules of $\cR$ satisfy the General Schema, then
  $\a_\cI$ is strongly normalizing.
\end{thm}

\begin{prf}
  One can easily prove that, for every term $u$ and computable
  substitution $\t$, $u\t$ is computable. In case where $u =
  f(\vec{u})$, we conclude by Lemma~\ref{lem-comp-fun-symb}. The
  theorem follows easily since the identity substitution is
  computable.
\end{prf}

~

It is possible to improve this termination result as follows. After
\cite{jouannaud97tcs}, if $\cR$ follows the General Schema and $\cR_1$
is a terminating set of non-duplicating\footnote{~No metavariable
  occurs more often in the right-hand side than in the left-hand
  side.}  first-order rewrite rules, then $\cR \cup \cR_1$ is also
terminating.



\section{Application of the General Schema to HRSs}
\label{sec-hrs}

We just recall what is a HRS. The reader can find precise definitions
in \cite{mayr98tcs}. A HRS $\cH$ is a pair $(\cA, \cR)$ made of a
HRS-alphabet $\cA$ and a set $\cR$ of HRS-rewrite rules over $\cA$. A
HRS-alphabet is a triple $(\cB, \cX, \cF)$ where $\cB$ is a set of
base types, $\cX$ is a family $(X_s)_{s\in T(\cB)}$ of variables and
$\cF$ is a family $(F_s)_{s\in T(\cB)}$ of function symbols. The
corresponding HRS-terms are the terms of the simply-typed
$\l$-calculus built over $\cX$ and $\cF$ that are in $\e$-long
$\b$-normal form.

So, a HRS $\cH$ can be seen as an IDTS $\ps{\cH}$ with the same
symbols, the arity of which being determined by the maximum number of
arguments they can take, plus the symbol $@$ for the application.
Hence it is a $\b$-IDTS. In \cite{oostrom93hoa}, van Oostrom and van
Raamsdonk studied this translation in detail and proved:


\begin{lem}[Van Oostrom and van Raamsdonk \cite{oostrom93hoa}]
  Let $\cH$ be a HRS. If $u \a_\cH v$ then $\cI(u) \a_{\cI(\cH)}
  \a_\b^* \cI(v)$ where $\cI(v)$ is in $\b$-normal form.
\end{lem}

\comment{Note that the proof is not easy since, in HRSs, the terms are always
in $\b$-normal form while, in IDTSs, rules may create $\b$-redexes.
Hence, it requires to show that a sequence of rewrites is equivalent
to a sequence of rewrites whose intermediate terms are
$\b$-normalized.}


As a consequence, $\cH$ is strongly normalizing if $\ps{\cH}$ so is.
Thus, the General Schema can be used on $\ps{\cH}$ for proving the
termination of $\cH$. In fact, it can be used directly on $\cH$ if we
adapt the notions of accessible subterm and computable closure to
HRSs. See Appendix~\ref{app-gs-hrs} for details.

\begin{thm}[Strong normalization for HRSs]
\label{thm-sn-hrs}
Let $\cH = (\cA, \cR)$ be a HRS satisfying the assumptions (A). If all
the rules of $\cR$ satisfy the General Schema for HRSs, then $\a_\cH$
is strongly normalizing.
\end{thm}

\begin{prf}
  This results from the fact proved in Appendix~\ref{app-gs-hrs} that,
  if $\cH$ follows the General Schema for HRSs then $\ps{\cH}$
  follows the General Schema for IDTSs.
\end{prf}



\section{Confluence of IDTSs}
\label{sec-confluence}

First of all, since an IDTS is a sub-CRS, it is confluent whenever the
underlying CRS is confluent. This is the case if it is weakly
orthogonal, \ie it is left-linear and all (higher-order) critical
pairs are equal \cite{oostrom94thesis}, or if it is left-linear and
all critical pairs are development closed \cite{oostrom95hoa}.

Now, one may wonder whether Nipkow's result for local confluence of
HRSs \cite{mayr98tcs} may be applied to IDTSs. To this end, we need to
interpret an IDTS as a HRS. This can be done in the following natural
way:


\newcommand{\aue}{\!\au^\e}

\begin{dfn}[Natural translation of IDTSs into HRSs]
  \hfill An IDTS-alphabet $\cA = (\cB, \cX, \cF, \cZ)$ can be
  naturally translated into the HRS-alphabet $\cH(\cA) = (\cB, \cX',
  \cF')$ where:
\begin{lst}{--}
\item $X'_{s_1\a \ldots\a s_n\a {\tt s}} = X_{s_1\a \ldots\a s_n\a
    {\tt s}} \cup \bigcup_{0\le p\le n} Z_{s_1, \ldots, s_p, s_{p+1}\a
    \ldots\a s_n\a {\tt s}}$
\item $F'_{s_1\a \ldots\a s_n\a {\tt s}} = \bigcup_{0\le p\le n}
  F_{s_1, \ldots, s_p, s_{p+1}\a \ldots\a s_n\a {\tt s}}$
\end{lst}

\noindent
An IDTS-metaterm $u$ is naturally translated into a HRS-term $\cH(u)$
as follows:

\begin{tabular}{cc}
\begin{minipage}{5cm}
\begin{lst}{--}
\item $\cH(x) = x \aue$
\item $\cH([x]u) = \l x. \cH(u)$
\end{lst}
\end{minipage}
&\begin{minipage}{5cm}
\begin{lst}{--}
\item $\cH(f(\vec{u})) = (f ~\cH(\vec{u})) \aue$
\item $\cH(Z(\vec{u})) = (Z ~\cH(\vec{u})) \aue$
\end{lst}
\end{minipage}
\end{tabular}

\noindent
Finally, an IDTS $\cI = (\cA, \cR)$ is translated into the HRS
$\cH(\cI) = (\cH(\cA), \cH(\cR))$ where $\cH(\cR) = \{ \cH(l) \a
\cH(r) ~|~ l\a r \in \cR \}$.
\end{dfn}

However, for Nipkow's result to hold, the rewrite rules must be of
base type, which is not necessarily the case for IDTSs. This is why,
in their study of the relations between CRSs and HRSs
\cite{oostrom93hoa}, van Oostrom and van Raamsdonk defined a
translation from CRSs to HRSs, also denoted by $\ps{~}$, which uses a
new symbol $\L$ for forcing the translated terms to be of base type.
Furthermore, they proved that (1) if $u \a_\cI v$ then $\ps{u}
\a_\ps{\cI} \ps{v}$, and (2) if $\ps{u} \a_\ps{\cI} v'$ then there is
a term $v$ such that $\ps{v} = v'$ and $u \a_\cI v$. In fact, it is no
more difficult to prove the same property for the translation $\cH$.
As a consequence, since $\ps{~}$ (resp. $\cH$) is injective, the
(local) confluence of $\ps{\cI}$ (resp.  $\cH(\cI)$) implies the
(local) confluence of $\cI$. Thus it is possible to deduce the local
confluence of $\cI$ from the analysis of the critical pairs of
$\ps{\cI}$ (resp. $\cH(\cI)$), and indeed, it turns out that
$\ps{\cI}$ and $\cH(\cI)$ have the ``same'' critical pairs (see the
proof of Theorem~\ref{thm-loc-confl} in Appendix~\ref{app-proofs} for
details). Identifying $\cI$ with its natural translation $\cH(\cI)$,
we claim that:

\begin{thm}
\label{thm-loc-confl}
If every critical pair of $\cI$ is confluent, then $\cI$ is locally
confluent.
\end{thm}

It could also have been possible to consider the translation $\cH'$
which is identical to $\cH$ but pulls down to base type the rewrite
rules by taking $\cH'(f(\vec{l}) \a r) = (f ~\cH(\vec{l}) ~\vec{x}) \a
v$ if $\cH(r) = \l \vec{x}. v$ with $v$ of base type. Note that the
left-hand side is still a pattern. Then, it is possible to prove that
$\cH(\cI)$ and $\cH'(\cI)$ have also the same critical pairs.



\section{Conclusion}

In Inductive Data Type Systems (IDTSs) \cite{blanqui98tcssub}, the use
of first-order matching does not allow to define some functions as
expected, resulting in non-confluent computations. By extending IDTS
with the higher-order pattern-matching mechanism of Klop's Combinatory
Reduction Systems (CRSs) \cite{klop93tcs}, we solved this problem and
made clear the relation between IDTSs and CRSs: IDTSs with
higher-order pattern-matching are simply-typed CRSs.

We extended a decidable termination criterion defined for IDTSs with
first-order matching and called the General Schema
\cite{blanqui98tcssub} to the case of higher-order pattern-matching,
and we proved that a rewrite system following this schema is
strongly-normalizing.

We also compared this unified approach to Nipkow's Higher-order
Rewrite Systems (HRSs) \cite{mayr98tcs}. First, we proved that the
extended General Schema can be applied to HRSs. Second, we show
how Nipkow's higher-order critical pair analysis technique for proving
local confluence can be applied to IDTSs.

~

Now, several extensions should be considered.

We did not take into account the interpretation defined in
\cite{blanqui98tcssub} for dealing with definitions over strictly
positive types (like Brouwer's ordinals or process algebra). However,
we expect that it can also be adapted to higher-order
pattern-matching.

It is also important to be able to relax the pattern condition which
says that metavariables must be applied to distinct bound variables.
But it is not clear how to prove the termination with Tait's
computability predicates technique when this condition is not
satisfied.

Another point is that some computations often need to be performed
within some equational theories like commutativity or commutativity
and associativity of some function symbols. It would be interesting to
know if the General Schema technique can be adapted for dealing with
such equational theories.

Finally, one may wonder whether all these results could be establish
in the more general framework of van Oostrom and van Raamsdonk's
Higher-Order Rewriting Systems (HORSs) \cite{oostrom94thesis,
  raamsdonk96thesis}, under some suitable conditions over the
substitution calculus.

~

\noindent
{\bf Acknowledgments:} I am very grateful to D. Kesner, A. Boudet and
J.-P. Jouannaud for their suggestions and remarks on previous versions
of this paper. I also thank the anonymous referees for their useful
comments.



\beginappendixes

\append{Relations between $\cI$ and $\b\cI$}
\label{app-beta-idts}

While the strong normalization of $\b\cI$ trivially implies the strong
normalization of $\cI$, it is an open problem whether the converse
holds. The difficulty comes from the fact that $\b$ may create
$\cI$-redexes and that $\cI$ may create $\b$-redexes.

In the case where $@$ is a symbol of $\cI$, the strong normalization
of $\cI$ does not imply the strong normalization of $\b\cI$, as
exemplified by the following counter-example due to Okada
\cite{okada89issac}. The non-left-linear rule

\begin{center}
$f(@(Z,Z'),Z') \a f(@(Z,Z'),@(Z,Z'))$
\end{center}

\noindent
terminates since each rewrite eliminates a $f$-redex ($@(Z,Z') \neq
Z'$), while its combination with $\b$ gives, by taking $Z = [x]x$, the
following infinite sequence of rewrites:

\begin{center}
  $f(@([x]x, y),y) \a f(@([x]x, y), @([x]x, y)) \a_\b f(@([x]x, y), y)
  \a \ldots$
\end{center}

In the case where all symbols are first-order, \ie all their arguments
are of base type, Breazu-Tannen and Gallier \cite{breazu89icalp,
  breazu91tcs} and Okada \cite{okada89issac} showed that it works.
Indeed, in this case, there cannot be interactions between rewriting
and $\b$-reduction.

~

Another problem is whether the confluence of $\cI$ implies the
confluence of $\b\cI$. This is not true in general even if $@$ is not
a symbol of $\cI$, as exemplified by a counter-example due to Klop
\cite{klop80thesis} using the non left-linear rule $f(x,x) \a a$.

On the other hand, it works when all function symbols are first-order
(even though the rules are not left-linear), as shown by the
pioneering work of Breazu-Tannen \cite{breazu88lics}.

With higher-order function symbols (\ie with arguments of functional
type), M\"uller proved in \cite{muller92ipl} that it works if the
rules are left-linear, contain no abstraction and no variable free in
the left-hand side is applied. In \cite{klop80thesis}, Klop showed
that it also works, this time with higher-order pattern-matching, when
$\cI$ is orthogonal, \ie the rules are left-linear and there is no
critical pair. Finally, van Oostrom \cite{oostrom94thesis} extended
these two results by proving that weakly orthogonal systems (systems
that are left-linear and whose critical pairs are equal) are
confluent.



\append{General Schema for HRSs}
\label{app-gs-hrs}


First of all, we precisely define the translation $\ps{~}$ from HRSs
to $\b$-IDTSs and the notions of accessible subterm and computable
closure for HRSs. Then, we prove that this notions are indeed
equivalent to the ones for IDTSs.

\begin{dfn}
  A HRS-alphabet $\cA = (\cB, \cX, \cF)$ is translated into the
  $\b$-extension $\b\cA'$ of the IDTS-alphabet $\cA' = (\cB, \cX,
  \cF', \cZ)$ where:
\begin{lst}{--}
\item $F'_{s_1, \ldots, s_n, {\tt s}} = F_{s_1 \a \ldots \a s_n \a
    {\tt s}}$,
\item $Z_{s_1, \ldots, s_n, {\tt s}} = \{ x \in X_{s_1\a \ldots\a s_n\a {\tt
      s}} \cap FV(l) ~|~ l \a r \in \cR \}$. 
\end{lst}

\noindent
A HRS-term $u$ is translated into an IDTS-term $\ps{u}$ as follows:

-- $\ps{\l x. u} = [x] \ps{u}$
\quad\quad -- $\ps{(x ~\vec{u})} = @(x, \ps{\vec{u}})$
\quad\quad -- $\ps{(f ~\vec{u})} = f(\ps{\vec{u}})$

\noindent
Assuming that bound variables are always taken away from the set $Z =
\{ x \in FV(l) ~|~ l \a r \in \cR \}$, a HRS-rewrite rule $l \a r$ is
translated into the IDTS-rewrite rule $\dps{l} \a \dps{r}$ where
$\dps{~}$ is defined as follows:\\
\begin{tabular}{cc}
\begin{minipage}{4cm}
\begin{lst}{--}
\item $\dps{\l x. u} = [x] \dps{u}$
\item $\dps{(f ~\vec{u})} = f(\dps{\vec{u}})$
\end{lst}
\end{minipage}
&\begin{minipage}{6cm}
\begin{lst}{--}
\item $\dps{(x ~\vec{u})} = @(x, \dps{\vec{u}})$ ~if $x \notin Z$
\item $\dps{(x ~\vec{u})} = x(\vec{u}\ad_\e)$ ~if $x \in Z$
\end{lst}
\end{minipage}
\end{tabular}

\noindent
Finally, a HRS $\cH = (\cA, \cR)$ is translated into the $\b$-IDTS
\,$\cI(\cH) = (\b\cA', \cR)$ where $\cR = \{ \dps{l} \a \dps{r} ~|~
l\a r \in \cR \} \cup \{ \b_{s,t} ~|~ s,t \in T(\cB) \}$. Moreover,
the constructors of a type ${\tt s}$ are the function symbols $c \in
F_{s_1, \ldots, s_n, {\tt s}}$ that are positive and not at the head
of a left-hand side of a rule of $\cR$.
\end{dfn}


\begin{dfn}[Accessible subterms for HRSs]
  The set $Acc'(v)$ of {\em accessible subterms} of a HRS-term $v$ is
  the smallest set such that:
\begin{enum}
\item $v \in Acc'(v)$
\item if $\l x. u \in Acc'(v)$, then $u \in Acc'(v)$
\item if $(c ~\vec{u}) \in Acc'(v)$ is of base type, then each $u_i
  \in Acc'(v)$
\item if $(f ~\vec{u}) \in Acc'(v)$ is of base type and $u_i$ is of
  basic type, then $u_i \in Acc'(v)$
\item if $(x ~\vec{u}) \in Acc'(v)$ is of base type and $x \notin
  FV(\vec{u}) \cup FV(v)$, then each $u_i \in Acc'(v)$
\end{enum}
\end{dfn}

We could have taken into account the case (5) of
Definition~\ref{dfn-acc} with the assertion: if $(u ~\vec{x}) \in
Acc'(v)$ is of base type and $\{ \vec{x} \} \cap (FV(u) \cup FV(v)) =
\emptyset$, then $u \in Acc'(v)$. But, in this case, $u$ must be a
variable. If it is a bound variable, then it is not useful. And if it
is a free variable, then it cannot be translated into an IDTS term.
This corresponds to the abuse of notation $Z \in Acc(v)$.

\begin{lem}
If $u \in Acc'(v)$ then $\dps{u} \in Acc(\dps{v})$.
\end{lem}

\begin{prf}
By induction on the definition of $Acc'(v)$.
\end{prf}


\begin{dfn}[Computable closure for HRSs]
  Given a function symbol $f \in F_{s_1\a \ldots\a s_n\a {\tt s}}$,
  the {\em computable closure} $\C'_f(\vec{l})$ of a HRS-term $(f
  ~\vec{l})$ is the least set $\C$ such that:

\begin{enum}
\item if $x \in FV(\vec{l}) \cap X_{t_1 \a \ldots \a t_p \a {\tt t}}$,
  $\vec{v}$ are $p$ terms $\e$-equivalent to distinct bound variables
  such that $(x ~\vec{v}) \in Acc'(\vec{l})$, and $\vec{u}$ are $p$
  terms of $\C$ of respective types $t_1, \ldots, t_p$, then $(x
  ~\vec{u}) \in \C$;
\item if $x \in X_{t_1 \a \ldots \a t_p \a {\tt t}} \setminus Z$ and
  $\vec{u}$ are $p$ terms of $\C$ of respective types $t_1, \ldots,
  t_p$, then $(x ~\vec{u}) \in \C$;
\item if $c \in C_{\tt t} \cap F_{t_1 \a \ldots \a t_p \a {\tt t}}$
  and $\vec{u}$ are $p$ terms of $\C$ of respective types $t_1,
  \ldots, t_p$, then $(c ~u_1 \ldots u_p) \in \C$;
\item if $u\in \C$ then $\l x. u \in \C$;
\item if $g \in F_{t_1 \a \ldots \a t_p \a {\tt t}}$, $g <_\cF f$ and
  $\vec{u}$ are $p$ terms of $\C$ of respective types $t_1, \ldots,
  t_p$, then $(g ~\vec{u}) \in \C$;
\item if $g \in F_{t_1 \a \ldots \a t_p \a t}$, $g =_\cF f$ and
  $\vec{u}$ are $p \ge 1$ terms of $\C$ of respective types $t_1,
  \ldots, t_p$ such that $\vec{u} ~\hle\statt \vec{l}$, then $(g
  ~\vec{u}) \in \C$.\footnote{~$\hle$ must of course be adapted to the
    HRS formalism.}
\end{enum}
\end{dfn}

We did not take into account the case (4) of
Definition~\ref{dfn-comp-closure} since we have to build terms in
$\b$-normal form.

\begin{lem}
  If $u \in \C'_f(\vec{l})$ then $\dps{u} \in \C_f(\dps{\vec{l}})$.
\end{lem}

\begin{prf}
By induction on the definition of $\C'_f(\vec{l})$.
\end{prf}


\begin{dfn}[General Schema for HRSs]
  A HRS-rewrite rule $(f ~\vec{l}) \a r$ follows the General Schema
  for HRSs GS$'$ if $r \in \C'_f(\vec{l})$.
\end{dfn}

\begin{lem}
\label{lem-GS-H-IH}
If $\cH$ follows GS$'$ then $\cI(\cH)$ follows GS.
\end{lem}



\append{Proofs}
\label{app-proofs}


\subsection*{Property~\ref{prop-acc}}

By induction on the proof that $u \in Acc(v)$. The only not
straightforward cases are (5) and (6). For case (5), by induction
hypothesis, $@(u,x)\s = @(u\s,x) \in Acc(v\s)$. Since $x$ is bound in
$v$, $x \notin FV(u\s) \cup FV(v\s)$. Hence, $u\s \in Acc(v\s)$. Case
(5) is treated in a similar way.


\subsection*{Property~\ref{prop-cov-ordering}}

\begin{prfenum}
\item $\hlt$ is stable by valuation since, for all $r<q$, $v|_{pr}$ is
  not headed by a metavariable.
\item Since, for all $r<q$, $v|_{pr}$ is not headed by an abstraction,
  $\hlt$ preserves free variables: if $u ~\hlt~ v$ then $FV(u) \sle
  FV(v)$. Hence $\hlt$ is stable by substitution.
\item Since, for all $r < p$, $v|_r$ is not headed by a defined
  symbol, no rewrite can take place above $v|_p$. Hence,
  covered-subterm steps can be postponed.
\end{prfenum}


\subsection*{Corollary~\ref{cor-compat-acc-comp}}

$Z \in Acc(l)$ means in fact that there are distinct bound variables
$\vec{x}$ such that $Z(\vec{x}) \in Acc(l)$. Now, if $Z(\vec{x})\s =
u$ then $\s(Z) = \ul(\vec{x}). u$ and, by Property~\ref{prop-acc}, $u
\in Acc(v)$. Let $\t$ be a computable substitution such that $dom(\t)
\sle \{ \vec{x} \}$.  $dom(\t) \cap FV(v) = \emptyset$ since $\vec{x}$
can always be taken away from $FV(v)$. Thus, by
Lemma~\ref{lem-compat-acc-comp}, $u\t$ is computable. Therefore,
$\s(Z)$ is computable.


\subsection*{Lemma~\ref{lem-commut}}

Since $a$ and $b$ are well-founded, $(a \cup\, b)^* = c^*$ with $c =
\bigcup_{k,l \ge 0}^{kl \neq 0} a^k b^l$. Since $ab \sle b^*a$, for
any $k$ and $l$, there is $m \ge 0$ such that $a^k b^l \sle b^m a^k$.
Hence, for any $k$, there is $m \ge 0$ such that $c^k \sle b^m a^n$
where $n$ is the number of $a$-steps in $c^k$. $m$ and $n$ are both
increasing with $k$ and are bounded since $a$ and $b$ are
well-founded, hence there is some $k_0$ such that $m$ and $n$ are
constant for all $k \ge k_0$. Therefore, the number of $a$-steps in
$c^k$ is finite and, hence, the number of $b$-steps too.


\subsection*{Lemma~\ref{lem-comp-clos-correct}}

In fact, we prove that, for any computable valuation $\s$ such that
$FV(cod(\s)) \cap FV(r) = \emptyset$, for any computable substitution
$\t$ such that $dom(\t) \sle FV(r)$ and for any $r \in \C_f(\vec{l})$,
$r\s\t = r\t\s$ is computable, by induction on $\C_f(\vec{l})$.

\begin{prfenum}
\item $r = Z(\vec{v})$ where $Z$ is a metavariable accessible in
  $\vec{l}$ and $\vec{v}$ are metaterms of $\C$. We first prove it for
  a special case and then for the general case.
  \begin{lstx}{}{\usecounter{enumii}%
     \renewcommand{\makelabel}{(\alph{enumii})}%
     \labelsep=-3mm\itemindent=3mm}
 \item $\vec{v}$ is a sequence of distinct bound variables, say
   $\vec{x}$. Without loss of generality, we can assume that $\s(Z) =
   \ul(\vec{x}). w$. Then, $r\s\t = w\t$. Since $\s$ is computable and
   $dom(\t) \sle \{\vec{x}\} = FV(r)$, $w\t$ is computable.
 \item $r\s\t$ is a $\b$-reduct of the term $@([\vec{x}]
   Z(\vec{x})\s\t, \vec{v}\s\t)$ where $\vec{x}$ are fresh distinct
   variables. By case (1a) and (5) below, $[\vec{x}]Z(\vec{x})\s\t$ is
   computable and since, by induction hypothesis, the terms in
   $\vec{v}\s\t$ are also computable, $r\s\t$ is computable.
  \end{lstx}
\item $r$ is a variable $x$. Then, $r\s\t = x\t$ is computable
  since $\t$ is computable.
\item $r = c(\vec{v})$ where $\vec{v}$ are metaterms of $\C$. Then,
  $c(\vec{v})\s\t = c(\vec{v}\s\t)$. By induction hypothesis, the
  terms in $\vec{v}\s\t$ are computable, hence $r\s\t$ is computable.
\item $r = @(v,w)$ where $v$ and $w$ are metaterms of $\C$. By
  induction hypothesis, $v\s\t$ and $w\s\t$ are computable, hence
  $r\s\t = @(v\s\t, w\s\t)$ is computable.
\item $r = [x]v$ where $v$ is a metaterm of $\C$.  Then, $r\s\t =
  [x]v\s\t$ and $r$ must have some functional type, say $s \a t$. Let
  $w$ be a computable term of type $s$. To prove that $@(r\s\t,w)$ is
  computable, it suffices to prove that its reduct $v\s\t'$ where $\t'
  = \t \{ x \to w \}$ is computable (see \cite{blanqui98tcssub}).
  Since $x$ can always be taken outside of $dom(\t)$ and
  $FV(cod(\t))$, $\t' = \t \uplus \{ x \to w \}$. Moreover, $dom(\t')
  \sle FV(v)$ and $\t'$ is computable. Hence, by induction hypothesis,
  $v\s\t'$ is computable and $r\s\t$ is computable.
\item $r = h(\vec{w})$ where $h <_\cF f$ and $\vec{w}$ are metaterms
  of $\C$. Then, $r\s\t = h(\vec{w}\s\t)$. By induction hypothesis,
  the terms in $\vec{w}\s\t$ are computable.  Hence, since $(h,
  \vec{w}\s\t) \clt (f, \vec{u})$, by assumption, $r\s\t$ is
  computable.
\item $r = g(\vec{v})$ where $g =_\cF f$ and $\vec{v}$ are metaterms
  of $\C$ such that $\vec{v} ~~\hle\statt~ \vec{l}$. Then, $r\s\t =
  g(\vec{v}\s\t)$. By induction hypothesis, the terms in $\vec{v}\s\t$
  are computable.  Now, since $\hle$ is stable by valuation and
  substitution, $\vec{v}\s\t ~~\hle\statt~\\ \vec{l}\s\t = \vec{l}\t\s
  = \vec{l}\s$ (the terms in $\vec{l}$ are closed). Hence, since $(g,
  \vec{v}\s\t) \clt (f, \vec{u})$, by assumption, $r\s\t$ is
  computable.
\end{prfenum}


\subsection*{Theorem~\ref{thm-loc-confl}}

We are going to show that there is a one-to-one correspondence between
the critical pairs of $\ps{\cI}$ and the critical pairs of $\cH(\cI)$.
The theorem follows easily.

But, first of all, we recall some definitions and results of
\cite{oostrom93hoa}.

~

\noindent
Van Oostrom and van Raamsdonk's translation: An IDTS-alphabet $\cA=
(\cB, \cX$, $\cF$, $\cZ)$ is translated into the HRS-alphabet
$\ps{\cA} = (\{o\}, \cX', \cF')$ where:
\begin{lst}{--}
\item $X'_o = \bigcup_{s\in T(\cB)} (X_s \cup Z_s)$
\item $X'_{o_n} = \bigcup_{s_1, \ldots, s_n \in T(\cB)} Z_{s_1,
    \ldots, s_n}$ ($n \ge 1$, $o_0 = o$ and $o_{n+1} = o\a o_n$)
\item $F'_{o_n} = \bigcup_{s_1, \ldots, s_n \in T(\cB)} F_{s_1,
    \ldots, s_n}$
\end{lst}

\noindent
An IDTS-metaterm $u$ is translated into a HRS-term $\ps{u}$ as
follows:\\
\begin{tabular}{cc}
\hspace*{1cm}
\begin{minipage}{4cm}
\begin{lst}{--}
\item $\ps{x} = x$
\item $\ps{[x]u} = (\L ~\l x. \ps{u})$
\end{lst}
\end{minipage}
&\begin{minipage}{6cm}
\begin{lst}{--}
\item $\ps{f(\vec{u})} = (f ~\ps{\vec{u}})$
\item $\ps{Z(\vec{u})} = (Z ~\ps{\vec{u}})$
\end{lst}
\end{minipage}
\end{tabular}

\noindent
Finally, an IDTS $\cI = (\cA, \cR)$ is translated into the HRS
$\ps{\cI} = (\ps{\cA}, \ps{\cR})$ where $\ps{\cR} = \{ \ps{l} \a
\ps{r} ~|~ l\a r \in \cR \}$.

~

Since HRS terms are $\l$-terms in $\b$-normal $\e$-long form, when we
apply a substitution $\t$ to a term $u$, the result of $u\t$ must be
$\b$-normalized. Following Nipkow's prefix notation, we denote $u\t
\!\ad_\b$ by $\t u$.

Given two left-hand sides of rule $l_1$ and $l_2$, there is a critical
pair between them at a position $p \in Pos(l_1)$ such that $l_1|_p$ is
not of the form $\l \vec{x}. (Z ~\vec{u})$ with $Z$ being a free
variable, if there is a substitution $\t$ such that $\t (l_1|_p) = \t
l_2$ and $FV(cod(\t)) \cap BV(l_1,p) = \emptyset$, $BV(l_1,p)$ being
the set of abstracted variables on the path from the root of $l_1$ to
$p$.

Given an IDTS valuation $\s$, $\ps{\s}$ (resp. $\cH(\s)$) denotes the
HRS substitution such that $Z\ps{\s} = \l \vec{x}. \ps{u}$ (resp.
$Z\cH(\s) = \l \vec{x}. \cH(u)$) whenever $\s(Z) = \ul(\vec{x}). u$.

~

Van Oostrom and van Raamsdonk proved that:

\begin{prfenum}
\item $\ps{u\s} = \ps{\s}\ps{u}$
\item If $l$ is a pattern such that $\t\ps{l} = \ps{u}$, then there is a
  valuation $\s$ such that $\ps{\s} = \t$.
\end{prfenum}

It is no more difficult to prove the same lemmas for $\cH$.

~

We now come to the proof that there is a one-to-one correspondence
between the critical pairs of $\ps{\cI}$ and the critical pairs of
$\cH(\cI)$.

Let $l_1$ and $l_2$ be two left-hand sides of rule of $\cI$. Assume
that there is a substitution $\t$ and a position $p \in Pos(\ps{l_1})$
such that $\t (\ps{l_1}|_p) = \t \ps{l_2}$. Without loss of
generality, we can assume that, for every variable $Z$, $Z\t$ is of
the form $\l \vec{x}. \ps{u}$. Then $\t (\ps{l_1}|_p)$ and $\t
\ps{l_2}$ are both of the form $\ps{u}$. Hence, by (2), there is a
valuation $\s$ such that $\ps{\s} = \t$. On the other hand, there is a
position $p' \in Pos(l_1)$ such that $\ps{l_1}|_p = \ps{l_1|_{p'}}$
and a position $p'' \in Pos(\cH(l_1))$ such that $\cH(l_1)|_{p''} =
\cH(l_1|_{p'})$. Thus, by injectivity of $\ps{~}$, $(l_1|_{p'})\s =
l_2$ and, by (1), $\cH(\s) (\cH(l_1)|_{p''}) = \cH(\s) \cH(l_2)$.

The other way around is proved in a similar way.


\end{document}